%% file: Honey_4.tex
\documentclass[twocolumn,english]{revtex4-1}
\usepackage[latin9]{inputenc}
\usepackage{babel}
\usepackage{mathtools}
\usepackage{amsmath}
\usepackage{amssymb}
\usepackage{graphicx}
\usepackage[unicode=true,
 bookmarks=false,
 breaklinks=false,pdfborder={0 0 1},backref=section,colorlinks=false]
 {hyperref}

\makeatletter
\usepackage{babel}

\usepackage{amsthm}\usepackage{amsfonts}\usepackage{times}\usepackage{color}\usepackage{url}
\include{PabMacros}

\def\tr#1{\textbf{T}(#1)}
\def\S{\mathcal{S}}

\makeatother

\begin{document}

\title{The Dirac equation as a quantum walk over the honeycomb and triangular
lattices}

\author{Pablo Arrighi}

\email{pablo.arrighi@univ-amu.fr}

\selectlanguage{english}%

\affiliation{Aix-Marseille Univ, Université de Toulon, CNRS, LIS, Marseille, France  and IXXI, Lyon, France}

\author{Giuseppe Di Molfetta}

\email{giuseppe.dimolfetta@lis-lab.fr}

\selectlanguage{english}%

\affiliation{Aix-Marseille Univ, Université de Toulon, CNRS, LIS, Marseille, France  and Departamento de F{\'{i}}sica Te{ó}rica and IFIC, Universidad
de Valencia-CSIC, Dr. Moliner 50, 46100-Burjassot, Spain}

\author{Iván Márquez-Martín}

\email{ivan.marquez@uv.es}

\selectlanguage{english}%

\affiliation{Aix-Marseille Univ, Université de Toulon, CNRS, LIS, Marseille, France  and Departamento de F{\'{i}}sica Te{ó}rica and IFIC, Universidad
de Valencia-CSIC, Dr. Moliner 50, 46100-Burjassot, Spain}

\author{Armando Pérez}

\email{armando.perez@uv.es}

\selectlanguage{english}%

\affiliation{Departamento de F{\'{i}}sica Te{ó}rica and IFIC, Universidad
de Valencia-CSIC,Dr. Moliner 50, 46100-Burjassot, Spain}

\date{\today}
\begin{abstract}
A discrete-time Quantum Walk (QW) is essentially an operator driving
the evolution of a single particle on the lattice, through local unitaries.
Some QWs admit a continuum limit, leading to well-known physics partial
differential equations, such as the Dirac equation. We show that these
simulation results need not rely on the grid: the Dirac equation in $(2+1)$--dimensions can
also be simulated, through local unitaries, on the honeycomb or the
triangular lattice, both of interest in the study of quantum propagation on the non-rectangular grids, as in graphene-like
materials. The latter, in particular, we argue, opens the door for a generalization
of the Dirac equation to arbitrary discrete surfaces.
\end{abstract}


\maketitle

\section{Introduction}

We will describe two novel discrete-time Quantum Walks (QW), one the honeycomb lattice, and other on the triangular lattice, whose continuum limit is the Dirac equation in $(2+1)$--dimensions. Let us put this result in context.

{\em Quantum walks.} QWs are dynamics having the following characteristics: \emph{(i)} the state space is restricted to the one particle sector (a.k.a. one `walker'); \emph{(ii)} spacetime is discrete; \emph{(iii)} the evolution is unitary; \emph{(iv)} the evolution is homogeneous, that is translation-invariant and time-independent, and \emph{(v)} causal (a.k.a. `non-signalling'), meaning that information propagates at a strictly bounded speed. Their study is blossoming, for two parallel reasons.\\ 
One reason is that a whole series of novel Quantum Computing algorithms, for the future Quantum Computers, have been discovered via QWs, e.g. \cite{BooleanEvalQW,ConductivityQW} and are better expressed using QWs. The Grover search has also been reformulated in this manner. In these QW-based algorithms, the walker usually explores a graph, which is encoding the instance of the problem. No continuum limit is taken.\\
The other reason is that a whole series of novel Quantum Simulation schemes, for the near-future Quantum simulation devices, have been discovered via QWs, and are better expressed as QWs \cite{Bialynicki-Birula, MeyerQLGI}. Recall that quantum simulation is what motivated Feynman to introduce the concept of Quantum Computing in the first place \cite{FeynmanQC}. Whilst an universal Quantum Computer remains out-of-reach experimentally, more special-purpose Quantum Simulation devices are seeing the light, whose architecture in fact often ressembles that of a QW \cite{WernerElectricQW,Sciarrino}. In these QW-based schemes, the walker propagates on the regular lattice, and a continuum limit is taken to show that this converges towards some well-known physics equation that one wishes to simulate. As an added bonus, QW-based schemes provide: 1/ stable numerical schemes, even for classical computers---thereby guaranteeing convergence as soon as they are consistent \cite{ArrighiDirac}; 2/ simple discrete toy models of the physical phenomena, that conserve most symmetries (unitarity, homogeneity, causality, sometimes even Lorentz-covariance \cite{arrighi2014discrete,DArianoLorentz}, perhaps even general covariance \cite{MolfettaDebbasch2014Curved, ArrighiGRDirac3D})---thereby providing playgrounds to discuss foundational questions in Physics \cite{LloydQG}. It seems that QW are unravelling as a new language to express quantum physical phenomena.\\
Whilst the present work is clearly within the latter trend, technically it borrows from the former. Indeed, the QW-based schemes that we will describe depart from the regular lattice, to go to the honeycomb and triangular grid---which opens the way for QW-based simulation schemes on trivalent graphs. 

{\em Motivations.} That quantum simulation schemes need not rely on the regular lattice grid is mathematically interesting---but there are numerous other motivations for this departure from the rectangular grid.  One is the hot topic of simulating/modeling many quantum condensed matter systems dynamics, driven by the usual high-binding Hamiltonian or by the Dirac-like Hamiltonian, for example in graphene, and within crystals in general \cite{neto2009electronic}. This work would establish a connection between such physical phenomena and QWs. Another hot topic is related to topological phases. QW on triangulations should allow us to model all sorts of topologies as simplicial complexes, and hopefully help predict their transport properties \cite{Kitagawa2010}. The fact that our Triangular QW converges to the Dirac equation shows that we have have the right prediction at least in the flat case. Yet another motivation for exploring non-flat geometries is General Relativity. In fact, two of the authors have already developped QW models of the curved spacetime Dirac equation \cite{MolfettaDebbasch2014Curved,ArrighiGRDirac3D,DebbaschWaves}. These were on the regular lattice, using a non-homogeneous coin to code for the spacetime-dependent metric. We wonder whether a QW on triangulations can also model the curved spacetime Dirac equation, using a homogeneous coin but a spacetime-dependent triangulation. This problem is reminiscent of the question of matter propagation in triangulated spacetime, as arising, \textit{e.g.}, in Loop Quantum Gravity \cite{RovelliDirac}. Here again, the fact that our Triangular QW converges towards the Dirac equation demonstrates that we have the right prediction at least in the triangulation-of-flat-space case. Finally, let us mention the work of two of the authors which models the massive Dirac equation as a Dirac QW on a cylinder \cite{MolfettaCylinder}. QW on triangulations should allow us to vary the geometry of this cylinder, so as to model richer fields with just the massless Dirac QW. 

{\em Related works.} %The quantum walk (QW) is the quantum analogue of the classical random walk. As in the case of random walks, QWs can appear either under its discrete-time \cite{Y.AharonovL.Davidovich1993} or continuous-time \cite{PhysRevA.58.915} form. We will concentrate here on discrete-time QWs, first considered by Grössing and Zeilinger \cite{grossing1988quantum} in 1988, as simple one-particle quantum cellular automata, and later popularized in the physics community in 1993, by Y. Aharonov \cite{Y.AharonovL.Davidovich1993}. Honeycomb lattices have been studied both in continuous and discrete QWs. 
The Grover quantum search algorithm has been expressed as a QW on the honeycomb lattice in \cite{Abal2010} (and also in \cite{Foulger2015} with continuous time). It has also been expressed as a QW on the triangular lattice \cite{Matsue2015,Abal2012}. Again for quantum algorithmic purposes, \cite{Karafyllidis2015} studies the possibility to use graphene nanoribbons to implement quantum gates. From the quantum simulation perspective, QWs on the triangular lattices have been used to explore transport in graphene structures \cite{Bougroura2016,chandrashekar2013two}, and they have also been used to explore topological phases \cite{Kitagawa2010}---but no actual continuum limit is taken in these works. To the best of our knowledge, the only work that does take a continuum limit of a discrete-time QW whilst departing from the regular lattice is \cite{sarkar2017effective}, where a Dirac-like hamiltonian is recovered. What we show is that the exact Dirac hamiltonian can be recovered, both in the honeycomb and the triangular lattices. That this can be done is somewhat surprising. Indeed, in \cite{Erba}, the authors conducted a thorough investigation of isotropic QW of coin dimension $2$ over arbitrary Caley graphs abelian groups, from which it follows that only the square lattice supports the Dirac equation. Our results circumvents this no-go theorem, whilst keeping things simple, by making use of two-dimensional spinors which lie on the edges shared by adjacent triangles, instead of lying on the triangles themselves. Thus means that, per triangle, there are three thus including an additional degree of freedom associated to these edges.

{\em Plan.} In order to start gently, Sec. \ref{sec:regular}, reexplains how the Dirac equation in $(2+1)$--dimensions can be simulated by a QW on the regular lattice. In Sec. \ref{sec:honeycomb}, we reexpress the $(2+1)$-dimensional Dirac Hamiltonian in terms of derivatives along arbitrary three $2\pi/3$--rotated axes $u_{i}$. We use this expression in order to simulate the Dirac equation with a QW on the honeycomb lattice. In Sec. \ref{sec:triangular}, we introduce a QW on the triangular lattice, which will turn out to be equivalent to that on the honeycomb lattice. In \ref{sec:conclusion} we provide a summary and some perspectives.

\section{On the regular lattice}\label{sec:regular}

\noindent In this section, we recall a now well-known  QW on the regular lattice with axis $x$, $y$ and spacing $\varepsilon$, which has the Dirac Equation in the continuum limit. It arises by operator-splitting \cite{FillionLorinBandrauk} the original, one-dimensional Dirac QW \cite{BenziSucci,Bialynicki-Birula,MeyerQLGI}.

A possible representation of this equation is (in units such as $\hbar=c=1$)
: 
\begin{align}
\ii\partial_{t}\ket{\psi} & =H_{D}\ket{\psi}\quad\textrm{with}\quad H_{D}=p_{x}\sigma_{x}+p_{y}\sigma_{y}+m\sigma_{z}\label{eq:Dirac2D}
\end{align}
the Dirac Hamiltonian, $\sigma_{i}$ ($i=1,2,3$) the Pauli matrices,
$p_{i}$ the momentum operator components and $m$ the particle mass.

To simulate the above dynamics on the lattice, we define a Hilbert
space $\mathcal{H}=\mathcal{H}_{l}\otimes\mathcal{H}_{s}$, where
$\mathcal{H}_{l}$ represents the space degrees of freedom and is
spanned by the basis states $\ket{x=\varepsilon l_{1},y=\varepsilon l_{2}}$
with $l_{1},l_{2}\in\mathbb{Z}$, whereas $\mathcal{H}_{s}=Span\{|s\rangle/s\in\{-1,1\}\}$
describes the internal (spin) configuration. When acting on $\mathcal{H}_{l}$,
the $p_{i}$'s are called quasimomentum operators (since they no longer
satisfy the canonical commutation rules with the position operators).
Still, the translation operators are given by $\tr{j,\varepsilon}=\exp(-\ii\varepsilon p_{j})$
and verify that 
\begin{equation}
\tr{1,\varepsilon}\ket{x,y}=\ket{x+\varepsilon,y},\,\,\,\,\,\tr{2,\varepsilon}\ket{x,y}=\ket{x,y+\varepsilon}.\nonumber 
\end{equation}
By analogy with these notations, we introduce the time evolution operator
as $\tr{0,\varepsilon}=\exp(-\ii\varepsilon H_{D})$. In this way,
the time evolution of a state $\ket{\psi(t)}$ is given by 
\begin{equation}
\ket{\psi(t+\varepsilon)}=\tr{0,\varepsilon}\ket{\psi(t)}=\exp(-\ii\varepsilon H_{D})\ket{\psi(t)}\label{eq:evolint}
\end{equation}
After substitution of Eq. (\ref{eq:Dirac2D}) into this definition,
and making use of the Lie-Trotter product formula (assuming that $\varepsilon$
is small) we arrive at: 
\begin{eqnarray}
\tr{0,\varepsilon} & \simeq & e^{-\ii\text{\ensuremath{\varepsilon}}m\sigma_{z}}e^{-\ii\text{\ensuremath{\varepsilon}}p_{x}\sigma_{x}}e^{-\ii\text{\ensuremath{\varepsilon}}p_{y}\sigma_{y}}\nonumber \\
 & = & e^{-\ii\text{\ensuremath{\varepsilon}}m\sigma_{z}}He^{-\ii\text{\ensuremath{\varepsilon}}p_{x}\sigma_{z}}HH_{1}e^{-\ii\text{\ensuremath{\varepsilon}}p_{y}\sigma_{z}}H_{1}^{\dagger},\nonumber 
\end{eqnarray}
since $\sigma_{x}=H\sigma_{z}H$ with $H$ the Hadamard gate, and
$\sigma_{y}=H_{1}\sigma_{z}H_{1}^{\dagger}$ with $H_{1}=\frac{1}{\sqrt{2}}\left(\begin{array}{cc}
\ii & 1\\
-\ii & 1
\end{array}\right)$. Using the definition of $\sigma_{z}$, we get: 
\begin{align}
\tr{0,\varepsilon} & \simeq C_{\varepsilon}HT_{1,\varepsilon}HH_{1}T_{2,\varepsilon}H_{1}^{\dagger}\label{eq:regularQW} \\
\textrm{with}\quad C_{\varepsilon} & =\exp\pa{-\ii\varepsilon m\sigma_{z}}\nonumber \\
\textrm{and}\quad T_{j,\varepsilon} & =\sum_{s\in\{-1,1\}}|s\rangle\bra{s}\tr{j,s\varepsilon}.\nonumber 
\end{align}
where the $T_{j,\varepsilon}$ matrices are partial shifts. This defines the Dirac QW, which is known to converge towards the Dirac equation in $(2+1)$-dimensions \cite{ArrighiDirac}. 
%As can be readly seen, it has a product form. Such `alternate quantum walks' have the advantage of using a two-dimensional coin-space instead of a four-dimensional coin-space: fewer resources are needed for their implementation \cite{McGettrick}. It is still just one QW, i.e. a translation-invariant causal unitary operator.

\section{On the honeycomb lattice}\label{sec:honeycomb}

\label{sec:honeycomb}

We now introduce a QW over the honeycomb lattice (Fig. \ref{Fig:Honeyandtriangles})
which we show has the Dirac equation as its continuum limit. The results
of this Section will also help us in the next Section, when we introduce
a QW over the triangular lattice. Our starting point is Eq. (\ref{eq:evolint}),
with $H_{D}$ as defined in Eq. (\ref{eq:Dirac2D}). The basic idea
is to rewrite this Hamiltonian using partial derivatives (that will
then turn into translations) along the three $(u_{i})$ vectors that
characterize nearest-neighbors in the hexagonal lattice, instead of
the $u_{x}$ and $u_{y}$ vectors that do so in the regular lattice.
The vectors $u_{i},\,\,\,i=0,1,2$ are given by 
\begin{equation}
u_{i}=\cos(i\frac{2\pi}{3})u_{x}+\sin(i\frac{2\pi}{3})u_{y},\label{eq:ui}
\end{equation}
with $u_{x}$ and $u_{y}$ the unit vectors along the $x$ and $y$
directions. In terms of momentum operators, 
\begin{equation}
\pi_{i}=\cos(i\frac{2\pi}{3})p_{x}+\sin(i\frac{2\pi}{3})p_{y}.\nonumber 
\end{equation}

\noindent We then look for three $2\times2$ matrices $\tau_{i}$
satisfying the following conditions: 
\begin{itemize}
\item[(C1)] Each of them has $\{-1,1\}$ as eigenvalues, i.e. there exists a unitary $U_{i}$
such that 
\begin{equation}
\tau_{i}=U_{i}^{\dagger}\sigma_{z}U_{i}.\nonumber 
\end{equation}

\item[(C2)] We impose that $\sum_{i=0}^{2}\tau_{i}\pi_{i}=p_{x}\sigma_{x}+p_{y}\sigma_{y}$,
i.e. the Dirac Hamiltonian adopts the form 
\begin{equation}
H_{D}=\sum_{i=0}^{2}\tau_{i}\pi_{i}+m\sigma_{z}.\nonumber 
\end{equation}

\end{itemize}
It was surprising to us that these conditions lead to unique $(\tau_{i})$
matrices, up to a sign: 
\begin{align}
\tau_{0} & =\frac{2}{3}\sigma_{x}+\xi\sigma_{z}\nonumber \\
\tau_{1} & =-\frac{1}{3}\sigma_{x}+\frac{\sqrt{3}}{3}\sigma_{y}+\xi\sigma_{z}\nonumber \\
\tau_{2} & =-\frac{1}{3}\sigma_{x}-\frac{\sqrt{3}}{3}\sigma_{y}+\xi\sigma_{z}.\nonumber 
\end{align}
with $\xi=\pm\frac{\sqrt{5}}{3}$. Let us choose $\xi=\frac{\sqrt{5}}{3}$,
and notice that 
\begin{equation}
\sum_{i}\tau_{i}=\frac{\sqrt{5}}{3}\sigma_{z}.\label{eq:tauisigmaz}
\end{equation}
Thus 
\begin{equation}
e^{-\ii\varepsilon H_{D}}=e^{-\ii\varepsilon\left(\sum_{i}\tau_{i}\pi_{i}+\frac{3}{\sqrt{5}}m\sum_{i}\tau_{i}\right)}.\nonumber 
\end{equation}
As before, we use the Lie-Trotter product formula and obtain: 
\begin{equation}
e^{-\ii\varepsilon\left(\sum_{i}\frac{3}{\sqrt{5}}m\tau_{i}+\tau_{i}\pi_{i})\right)}\simeq\prod_{i=0}^{2}e^{-\ii\varepsilon\frac{3}{\sqrt{5}}m\tau_{i}}e^{-\ii\varepsilon\tau_{i}\pi_{i}}.\label{eq:HoneyTrotter}
\end{equation}
We now make use of condition (C1) to rewrite, for each $i$, 
\begin{equation}
e^{-\ii\varepsilon\tau_{i}\pi_{i}}=e^{-\ii\varepsilon U_{i}^{\dagger}\sigma_{z}U_{i}\pi_{i}}=U_{i}^{\dagger}e^{-\ii\varepsilon\sigma_{z}\pi_{i}}U_{i}=U_{i}^{\dagger}T_{i,\varepsilon}U_{i}\nonumber 
\end{equation}
where now the partial shifts $T_{i,\varepsilon}$ are defined through
the $\pi_{i}$ operators, instead of $p_{x}$ and $p_{y}$. Similarly,
for all $i$, 
\begin{equation}
e^{-\ii\varepsilon\frac{3}{\sqrt{5}}m\tau_{i}}=U_{i}^{\dagger}e^{-\ii\varepsilon\frac{3}{\sqrt{5}}m\sigma_{z}}U_{i}.\nonumber 
\end{equation}
Let $M=e^{-\ii\varepsilon\frac{3}{\sqrt{5}}m\sigma_{z}}$. Wrapping
it up, we have obtained a QW over the honeycomb lattice: 
\begin{equation}
\ket{\psi(t+\varepsilon)}=\left(\prod_{i=0}^{2}U_{i}^{\dagger}MT_{i,\varepsilon}U_{i}\right)\ket{\psi(t)}.\label{eq:honeyQW}
\end{equation}
which, by construction, has the Dirac Eq. \eqref{eq:Dirac2D} as its
continuum limit as $\varepsilon\rightarrow0$. By mere associativity
the QW rewrites as 
\begin{equation}
U_{0}\ket{\psi(t+\varepsilon)}=\left(\prod_{i=0}^{2}U_{i+1}U_{i}^{\dagger}MT_{i,\varepsilon}\right)U_{0}\ket{\psi(t)},\nonumber 
\end{equation}
Thus, if the matrix products $U_{i+1}U_{i}^{\dagger}$ could be made
independent of $i$  (with $i+1$ understood modulo $3$),
the QW could be reformulated to have a constant coin operator. Surprisingly,
this can be done thanks to a natural choice of the $U_{i}$ matrices,
expressed in terms of well-chosen rotations in the Bloch sphere, understood
as the set of possible spin operators. The natural choice for $U_{0}$
is $\mathcal{R}_{\sigma_{y}}(\alpha)=e^{-\ii\alpha\sigma_{y}/2}$,
the rotation of angle $\alpha=\arccos{\frac{\sqrt{5}}{3}}$ around
$\sigma_{y}$. Indeed $\mathcal{R}_{\sigma_{y}}(\alpha)$, maps the
Bloch vector of $\tau_{0}$ into the Bloch vector of $\sigma_{z}$:
\begin{equation}
\sigma_{z}=\mathcal{R}_{\sigma_{y}}(\alpha)\tau_{0}\mathcal{R}_{\sigma_{y}}^{\dagger}(\alpha).
\end{equation}
Next, we observe that the Bloch vectors $\tau_{i}$ are related by
a rotation of angle $2\pi/3$ around $\sigma_{z}$. For reasons that
will become apparent, it matters to us that the cube of this rotation
is the identity, which is obviously not the case for $\mathcal{R}_{\sigma_{z}}(\frac{2\pi}{3})=e^{-i\pi/3\sigma_{z}}$,
since it represents a spin 1/2 rotation and will acquire a minus sign
when applied three times. Hence we take 
$\S = e^{\ii\frac{\pi}{3}}\mathcal{R}_{\sigma_{z}}(\frac{2\pi}{3})$
instead. Then, the natural choice for
the matrices $U_{1}$ and $U_{2}$ is: 
\begin{equation}
U_{1} =U_{0}\S\qquad
U_{2} =U_{1}.\S\nonumber 
\end{equation}
Indeed, these again fulfill (C1): first the $\S$
unitary brings $\tau_{i}$ to $\tau_{0}$, and then the $U_{0}$ rotation
brings $\tau_{0}$ to $\sigma_{z}$. Now, the fact that the $U_{i}$
matrices are related by a unitary which cubes to the identity entails
that the products 
$U_{i+1}U_{i}^{\dagger}=U_{0}\S U_{0}^{\dagger}$
are independent of $i$. We introduce 
\begin{equation}
W=U_{0}\S U_{0}^{\dagger}M.\label{eq:W}
\end{equation}
Then, if we redefine the field up to an encoding, via 
\begin{equation}
\ket{\widetilde{\psi}(t)}\equiv U_{0}\ket{\psi(t)},\nonumber
\end{equation}
Then the Honeycomb QW rewrites as just: 
\begin{equation}
\ket{\widetilde{\psi}(t+\varepsilon)}=\left(WT_{2,\varepsilon}WT_{1,\varepsilon}WT_{0,\varepsilon}\right)\ket{\widetilde{\psi}(t)}.\label{eq:HoneyQW2}
\end{equation}
In other works, the Honeycomb QW just shifts the $\pm$-components
along $\pm u_{0}$, applies the fixed $U(2)$ matrix $W$ at each
lattice point, shifts the $\pm$-components along $\pm u_{1}$, applies
$W$ again, etc. For certain architectures it could well be that the
time homogeneity of the coins makes the scheme easier to implement
experimentally, compared to earlier alternate QW on the regular lattice
\cite{ArrighiDirac}.

\section{On the triangular lattice}\label{sec:triangular}

\begin{figure}
\includegraphics[width=0.5\columnwidth]{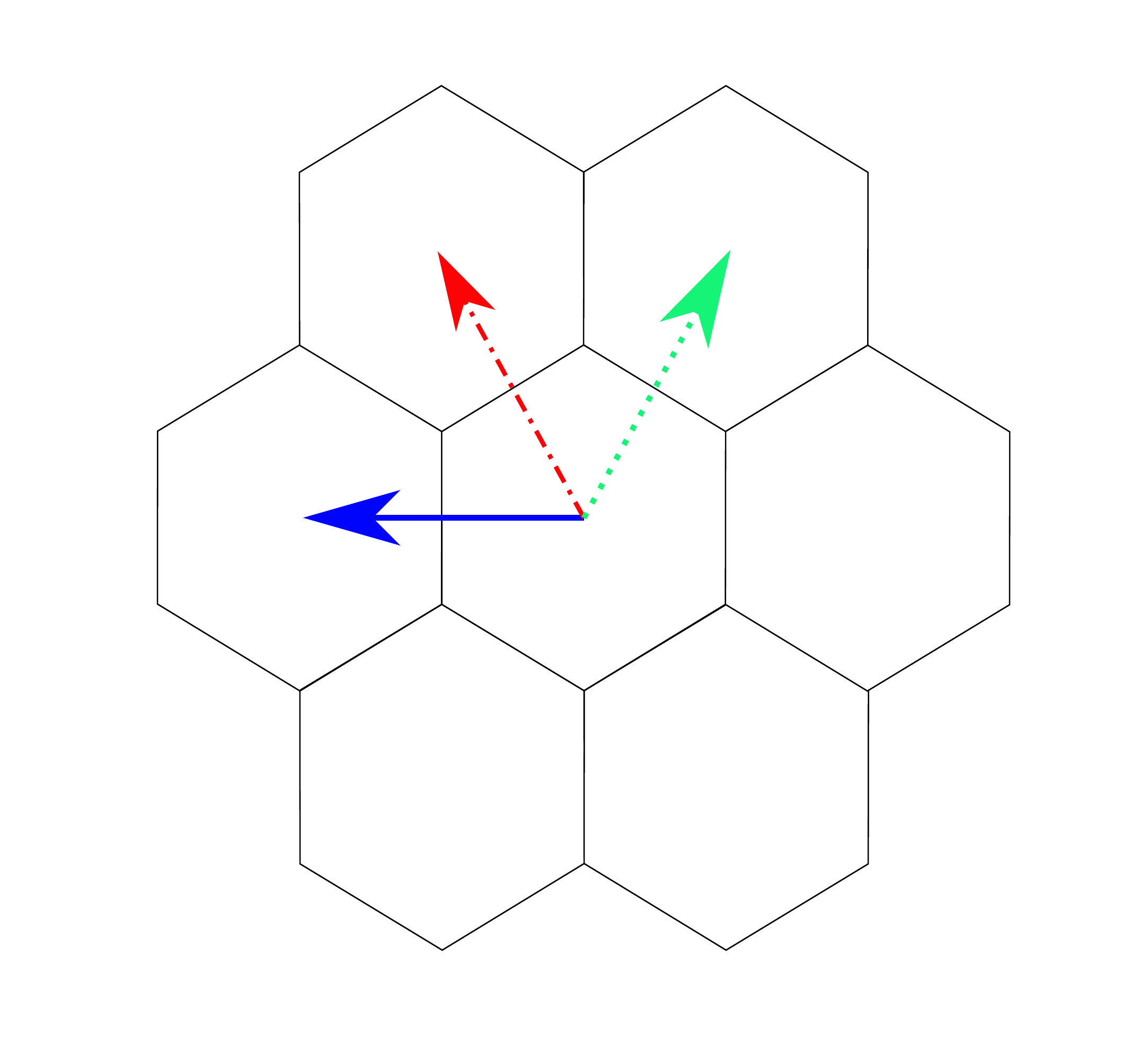}\includegraphics[width=0.5\columnwidth]{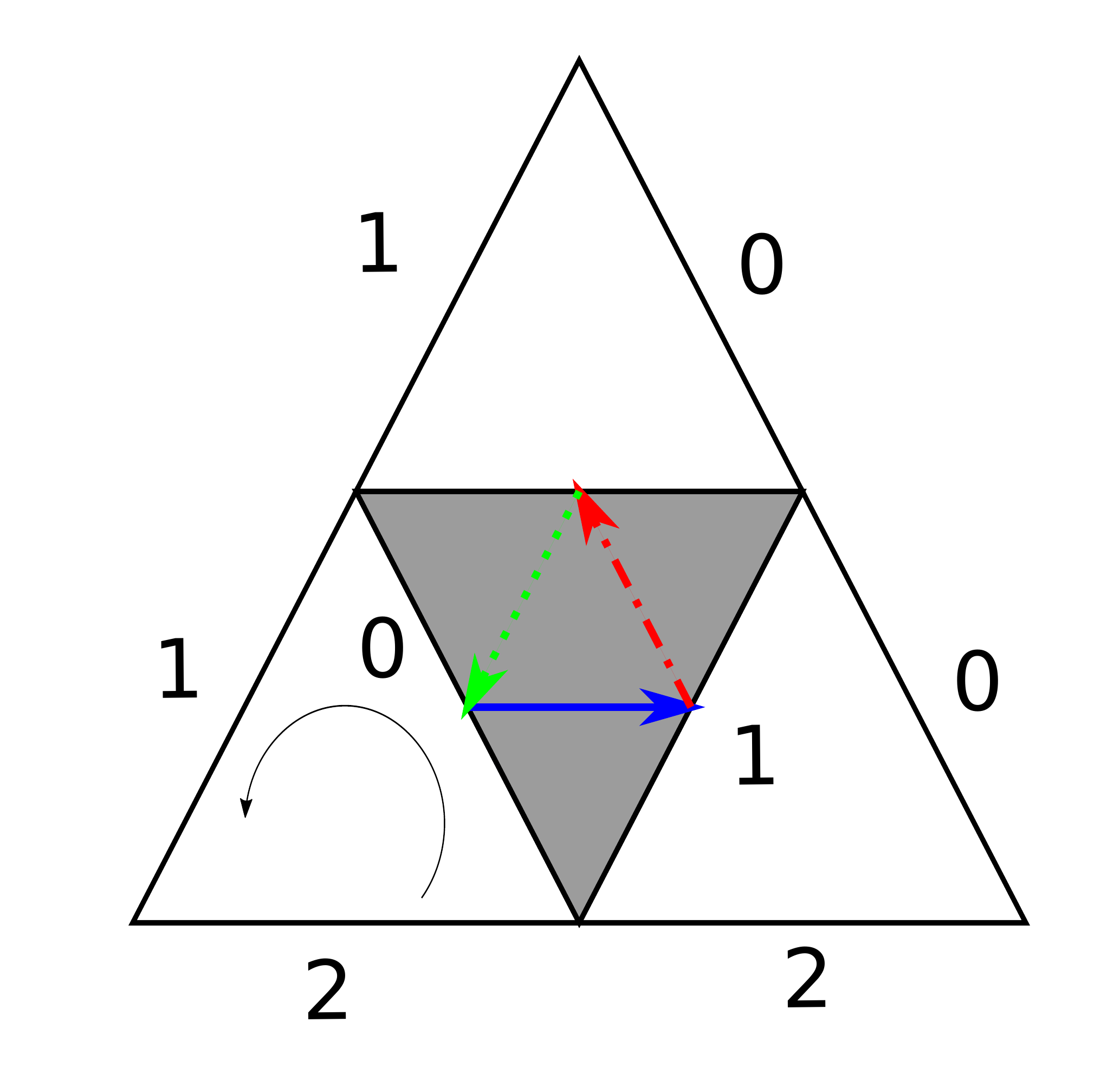}

\caption{(color online) Left: The Honeycomb QW. The particle moves first along the $u_{0}$
direction (blue solid line), then $u_{1}$ (red dot-dashed line) and finally $u_{2}$
(green dot line). Right: The Triangular QW. Starting at the edge $k=0$,
the dynamics is equivalent to the honeycomb QW, in three time-steps.
The circle line represents the counter-clockwise rotation operator. }

\label{Fig:Honeyandtriangles} 
\end{figure}

\noindent Having understood how to obtain the Dirac Eq. over the honeycomb
lattice will make it much easier to tackle the triangular or related lattice such as the kagome lattice \cite{ye2018massive}.
Let us first describe the lattice and its state space. Our triangles
are equilateral with sides $k=0,1,2$, see Fig. \ref{Fig:Honeyandtriangles}.
Albeit the drawing shows white and gray triangles, these differ only
by the way in which they were laid---they have the same orientation
for instance. Our two-dimensional spinors lie on the edges shared
by neighboring triangles. We label them $\psi(t,v,k)=\left(\begin{array}{c}
\psi^{\uparrow}(t,v,k)\\
\psi^{\downarrow}(t,v,k)
\end{array}\right)$, with $v$ a triangle and $k$ a side. But, since each spinor lies
on an edge, we can get to it from two triangles. For instance if triangle
$v_{0}$ (white) and $v_{1}$ (grey) are glued along their $k=1$
side, then $\psi(t,v_{0},1)=\psi(t,v_{1},1)$. In fact let us take
the convention that the upper (resp. lower) component of the spinor,
namely $\psi^{\uparrow}$ (resp. $\psi^{\downarrow}$), lies on the
white (resp. gray) triangle's side. From this perspective each triangle
hosts a $\mathbb{C}^{3}$ vector, e.g. $\psi(t,v_{0})=(\psi^{\uparrow}(t,v_{0},k))_{k=0\ldots2}^{T}$
and $\psi(t,v_{1})=(\psi^{\downarrow}(t,v_{1},k))_{k=0\ldots2}^{T}$.\\
 The dynamics of the Triangular QW is the composition of two operators.
The first operator, $R$, simply rotates every triangle anti-clockwise.
Phrased in terms of the hosted $\mathbb{C}^{3}$ vectors, the component
at side $k$ hops to side $(k+1\mod3)$. For instance $R\psi(t,v_{0})=(\psi^{\uparrow}(t,v_{0},k-1))_{k=2,0,1}$.
The second operator is just the application of the $2\times2$ unitary
matrix $W$ given in \eqref{eq:W}, to every two-dimensional spinor
of every edge shared by two neighboring triangles. Again we work on
pre-encoded spinors 
\begin{equation}
\widetilde{\psi}(t,v,k)=U_{k}\psi(t,v,k) \label{eq:encoding}
\end{equation}
where the $U_{k}$ are those of Sec. \ref{sec:honeycomb}, but this time
the chosen encoding depends on side $k$. Altogether, the Triangular
QW dynamics is given by: 
\begin{equation}
\left(\begin{array}{c}
\widetilde{\psi}^{\uparrow}(t+\varepsilon,v,k)\\
\widetilde{\psi}^{\downarrow}(t+\varepsilon,v,k)
\end{array}\right)=W\left(\begin{array}{c}
\widetilde{\psi}^{\uparrow}(t,v,k-1)\\
\widetilde{\psi}^{\downarrow}(t,e(v,k),k-1)
\end{array}\right)
\label{eq:triangulatQW}
\end{equation}
where $e(v,k)$ is the neighbor of triangle $v$ alongside $k$.

This Triangular QW is actually implementing the Honeycomb QW in a
covert way. Indeed, whereas the Honeycomb QW propagates the walker
along the three directions successively, the Triangular QW propagates
the walker along the three translation simultaneously---depending
on the edge at which it currently lies. Thus the walker will start
moving along one of the three direction depending on its starting
point, then another, etc. For instance, focusing on what happens to
spinors on edges $k=0$, we readily get 
\begin{equation}
\left(\begin{array}{c}
\widetilde{\psi}^{\uparrow}(\varepsilon,v,1)\nonumber\\
\widetilde{\psi}^{\downarrow}(\varepsilon,v,1)
\end{array}\right)=VM\left(\begin{array}{c}
\widetilde{\psi}^{\uparrow}(0,v,0)\\
\widetilde{\psi}^{\downarrow}(0,e(v,2),0)
\end{array}\right),
\end{equation}
which is equivalent to a translation along $u_{0}$ (as is clear from
Fig. \ref{Fig:Honeyandtriangles}), followed by the action of $W$.
But the result now lies on edges $k=1$, and will undergo a translation
along $u_{1}$ followed by the action of $W$, etc.

As a sanity check we computed the continuum limit obtained by letting
$\varepsilon\rightarrow0$ after three iterations of Eq. (\ref{eq:triangulatQW}).
The $0^{th}$ order is trivial. The $1^{st}$ is what defines the
dynamics. Let us align the middle of the side side $1$ of triangle
$v$ with the origin of the Euclidean space, so that $\psi(0,v,1)=\psi(0,0,0)$
in Cartesian coordinates. Expand the initial condition $\psi(0,x,y)$
as: 
\begin{equation}
\psi(0,x,y)=\psi(0,0,0)+\varepsilon x\partial_{x}\psi(0,0,0)+\varepsilon y\partial_{y}\psi(0,0,0)\nonumber
\end{equation}
where $x$ and $y$ are the coordinates in the lattice. As usual we
also expand the $M$ inside the $W$ as $\mathbb{I}-3\ii\varepsilon m\sigma_{z}/\sqrt{5}$.
After three steps of the Triangular QW we obtain (with a the help
of a computer algebra system): 
\begin{align}
\tr{0,3\varepsilon}\psi=\psi(0,0)-\frac{\sqrt{3}}{2}\varepsilon\left(\sigma_{x}\partial_{x}+\sigma_{y}\partial_{y}\right)\psi(0,0)\nonumber \\
-3 \ii\varepsilon m\sigma_{z}\psi(0,0)+\mathcal{O}(\varepsilon^{2})\nonumber
\end{align}
Using that $\tr{0,3\varepsilon}=\psi(0,0)+3\varepsilon\partial_{t}\psi(0,0)+\mathcal{O}(\varepsilon^{2})$,
and taking the limit $\varepsilon\rightarrow0$, we arrive to the
Dirac equation under the following form: 
\begin{align}
\ii\partial_{t}\psi(0,0)=\frac{\sqrt{3}}{6}\left(p_{x}\sigma_{x}+\sigma_{y}p_{y}\right)\psi(0,0)
+m\sigma_{z}\psi(0,0)\nonumber
\end{align}
The factor $\frac{\sqrt{3}}{6}$ comes from two reasons: the fact that continuous limit results from three-time steps and the fact that the distance between the middles of the sides of a triangle is $\frac{\sqrt{3}}{2}$. To get rid of this factor, it suffices to rescale the length of the spatial coordinates of the triangles by the same factor, or conversely to rescale time as $t'=\frac{6}{\sqrt{3}}t$.

\section{Summary and perspectives}\label{sec:conclusion}

{\em Summary.} We constructed a $2\times 2$ unitary $W$, defined in Eq. \eqref{eq:W}, which serves as the `coin' for both the Honeycomb QW and the Triangular QW. On the honeycomb lattice, each hexagon carries a $\mathbb{C}^2$ spin. The Honeycomb QW, defined in Eq. \eqref{eq:HoneyQW2}, simply alternates a partial shift along along the $u_i$-direction of \eqref{eq:ui}, followed by a $W$ on each hexagon, for $i=0,1,2$. On the triangular lattice, each side of each triangle carries a $\mathbb{C}$, so that each edge shared by two neighbouring triangles carries a $\mathbb{C}^2$ spin. The Triangular QW, defined in \eqref{eq:triangulatQW}, simply alternates a rotation of each triangle, and the application of $W$ at each edge. The simplicity of these QW-based schemes, compared to those of the regular lattice \eqref{eq:regularQW}, makes them not only elegant, but also easy to implement. Our main result states that, up to a simple, local unitary encoding given by \eqref{eq:encoding}, both the Honeycomb QW and the Triangular QW admit, as their continuum limit, the Dirac Eq. in $(2+1)$--dimensions.\\
{\em Perspectives.} Thus we have shown that such quantum simulations results need not rely on the grid. We believe that this constitutes an important step towards : modelling propagation in crytalline materials; identifying substrates for QW implementations; studying topological phases; understanding propagation in discretized curved spacetime; coding fields in closed dimensions. In the near future, we wish run numerical simulations, and to understand what happens when deforming the triangles, and whether similar results can be achieved in $(3+1)$--dimensions.\medskip\\ 
\noindent 
{\bf Update.} We recently became aware that another,  French-Australian, team was tackling the same problem. We agreed to swap papers a few days before arXiv submission, so that the two works would be independent, and yet cite each other. Manuscript \cite{Manuscript} is indeed very recommendable, as it goes further in terms of applications: electromagnetic field; gauge-invariance; numerical simulations. Their triangular walk is, however, an alternation of three different steps, that use different coins---whereas the present paper just iterates the very same step. This is both mathematically more elegant, and easier to implement. Thus the two works have turned out nicely complementary.\medskip\\
\noindent {\bf Acknowledgements.} We acknowledge an interesting discussion with M. C. Ba\~nuls. This work has been funded by the ANR-12-BS02-007-01 TARMAC grant, the STICAmSud project 16STIC05 FoQCoSS and the Spanish Ministerio
de Economía, Industria y Competitividad , MINECO-FEDER project FPA2017-84543-P,
SEV-2014-0398 and Generalitat Valenciana grant GVPROMETEOII2014-087, the project INFINITI CNRS. 

\bibliographystyle{plain}
\bibliography{honeyref}

\end{document}

%% file: PabMacros.tex
\newcommand{\couic}[1]{}
\newcommand{\couicfootnote}[1]{}
\newcommand{\couicefootnote}[1]{}

\newcommand{\ket}[1]{| #1 \rangle}
\newcommand{\bra}[1]{\langle #1 |}

\newcommand{\ii}{\mathrm i}

\newcommand{\pa}[1]{\left(#1\right)}